\documentclass[aps,prl,nobibnotes,nofootinbib,showpacs,preprint]{revtex4-1}

\usepackage{graphicx}
\usepackage{amsmath} 	

\newcommand{\ket}[1]{|#1\rangle}
\newcommand{\bra}[1]{\langle#1|}
\newcommand{\Fig}[1]{figure~\ref{#1}}

\newcommand{\Eq}[1]{equation~(\ref{#1})} 
 
\newcommand{\Eqsx}[3]{equations~(\ref{#1}),~(\ref{#2}),~and~(\ref{#3})} 
\newcommand{\w}{\omega}

\begin{document}

\title{Two-color Multiphoton Emission from Nanotips}

%\author{Wayne~Cheng-Wei~Huang}
%\address{Department of Physics and Astronomy, University of Nebraska-Lincoln, Lincoln, Nebraska 68588, USA}
%
%\author{Maria Becker}
%\address{Department of Physics and Astronomy, University of Nebraska-Lincoln, Lincoln, Nebraska 68588, USA}
%
%\author{Joshua Beck}
%\address{Department of Physics and Astronomy, University of Nebraska-Lincoln, Lincoln, Nebraska 68588, USA}
%
%\author{Herman~Batelaan}
%\ead{hbatelaan2@unl.edu}
%\address{Department of Physics and Astronomy, University of Nebraska-Lincoln, Lincoln, Nebraska 68588, USA}

\author{Wayne~Cheng-Wei~Huang}
\author{Maria Becker}
\author{Joshua Beck}
\author{Herman~Batelaan}
\email{email: hbatelaan2@unl.edu}

\affiliation{Department of Physics and Astronomy, University of Nebraska-Lincoln, Lincoln, Nebraska 68588, USA}

\begin{abstract}

Two-color multiphoton emission from polycrystalline tungsten nanotips has been demonstrated using two-color laser fields. The two-color photoemission is assisted by a three-photon multicolor quantum channel, which leads to a twofold increase in quantum efficiency. Weak-field control of two-color multiphoton emission was achieved by changing the efficiency of the quantum channel with pulse delay. The result of this study complements two-color tunneling photoemission in strong fields, and has potential applications for nanowire-based photonic devices. Moreover, the demonstrated two-color multiphoton emission may be important for realizing ultrafast spin-polarized electron sources via optically injected spin current.

\end{abstract}

 %maximum 4 pacs code, the first is the principal pacs code

\maketitle 

\section{Introduction}

Electron photoemission has played an important role in the advancement of ultrafast science \cite{Corkum, Krausz}. Recent studies of photoelectrons demonstrate the feasibility of using tip-like nanostructures as ultrafast light detectors and ultrafast electron sources \cite{Wimmer, Piglosiewicz, Kruger2011}. In these studies, carrier-envelope phase was used for ultrafast control of tunneling photoemission in strong fields. However, such methods are not effective in the weak-field regime, as photoemission in weak fields is accomplished through multiphoton processes \cite{Kruger2012}. The perturbative nature of such processes makes photoemission insensitive to the instantaneous field and the carrier-envelope phase. The weak-field regime is especially important for nanotip photoemission because in this regime high repetition rates are easily accessible, which can lead to bright photoemission electron sources while avoiding laser induced damage. 

In this report, we show weak-field control of two-color photoemission from a nanotip by opening a multicolor quantum channel. In the strong-field regime, two-color photoemission is controlled by the asymmetric waveform of a two-color field which facilitates directional tunneling \cite{Betsch, Shafir, Kim, Mancuso}. In the weak-field regime, two-color photoemission control will, however, be based on opening and closing a multicolor quantum channel for multiphoton emission. Multicolor quantum channels are multiphoton transitions in which photons of different colors are simultaneously absorbed or emitted \cite{Mahou}. Given a fixed photon flux, a multicolor quantum channel can be used as a valve to control the output photocurrent.

The opening of a multicolor quantum channel can lead to a twofold increase in quantum efficiency. The multicolor quantum channel and the associated increase in quantum efficiency have potential applications for nanowire-based photonic devices \cite{Bulgarini, Mustonen, Swanwick}. With the appropriate work function and laser wavelengths, ultrafast control in weak fields may be obtained through quantum interference between single-color and multicolor quantum channels \cite{Yin, Schafer, Ray}. The demonstrated two-color multiphoton emission may also provide a pathway for realizing ultrafast spin-polarized electron sources via optically injected spin current \cite{Bhat, Hubner, Stevens}. 

\begin{figure}[t]
\centering
\scalebox{0.35}{\includegraphics{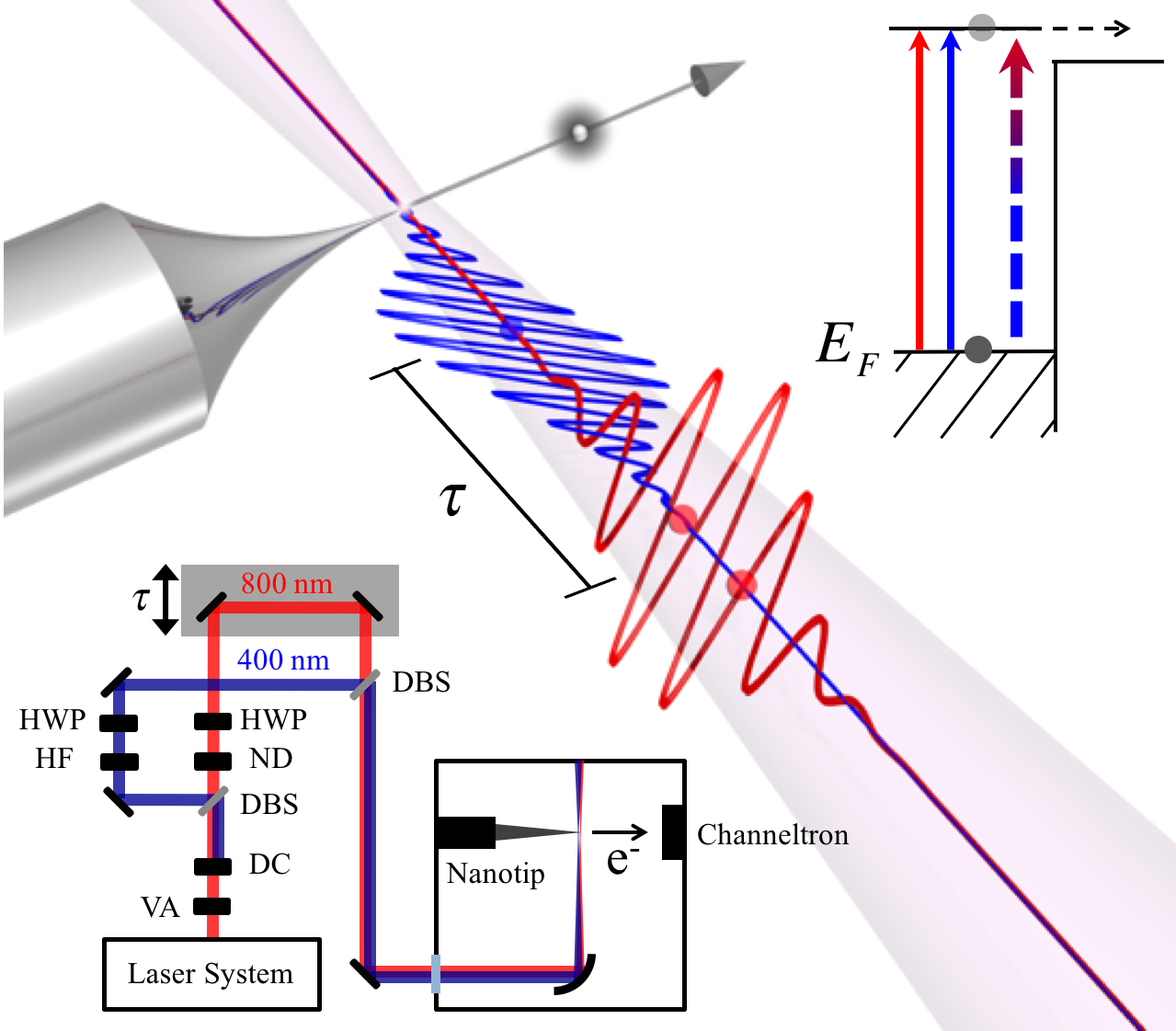}}
\caption{Schematic of the experimental setup. A tungsten nanotip is irradiated by two-color fields. The two-color multiphoton emission is assisted by the multicolor quantum channel. Top right: multicolor quantum channels (dashed arrow) open up as the two-color pulses overlap. Bottom left: a Mach-Zehnder interferometer controls the delay $\tau$ of the 800 $nm$ pulse. Acronyms VA, DC, DBS, HF, ND, and HWP stand for variable attenuator, frequency-doubling crystal, dichroic beamsplitter, high-pass filter, neutral density filter, and half-waveplate.}
\label{fig:tip_field}
\end{figure}

\section{Experimental Setup}

\begin{figure}[t]
\centering
\scalebox{0.5}{\includegraphics{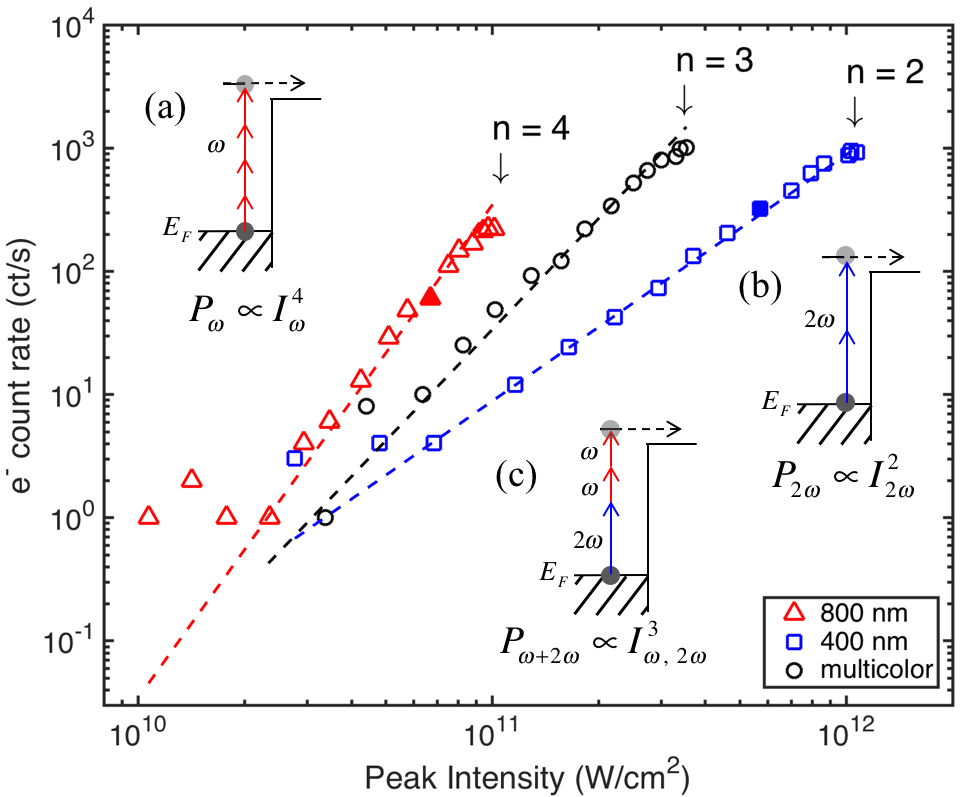}}
\caption{Intensity dependence of single-color and two-color multiphoton emission. The single-color signals are plotted against single-color field intensities, $I_{\w}$ and $I_{2\w}$. To compare emission curves for different wavelengths, single-color data are shifted by scaling the intensities to $0.1 \times I_{\w}$ and $25 \times I_{2\w}$. The multicolor signal is background-subtracted and plotted against the two-color field intensity $I_{\w,2\w} \equiv (I^2_{\w}I_{2\w})^{1/3}$. Dashed lines with slopes $n =$ 4, 3, and 2 are guides to the eye. Insets (a), (b), and (c) show the corresponding diagrams for the 800 $nm$ four-photon, the 400 $nm$ two-photon, and the two-color three-photon processes described in \Eqsx{single-color_probw}{single-color_prob2w}{multicolor_prob}.}
\label{fig:intensity}
\end{figure}

Tungsten nanotips were used in our experiment because they are robust photoelectron emitters. The tips were prepared by electrochemically etching a polycrystalline tungsten wire. Tip radii were estimated to be around 50 $nm$ \cite{Barwick}. A schematic of the experimental setup is given in \Fig{fig:tip_field}. A linearly-polarized, 400 $nm$ pulse was generated collinearly from a linearly-polarized, 800 $nm$ pulse using a frequency-doubling crystal (BBO Type I, thickness 0.5 $mm$). The 800 $nm$ pulse was provided by an amplified laser. The two pulses were separated by a dichroic beamsplitter as they entered a Mach-Zehnder interferometer. A high-pass filter was placed in the optical arm of the 400 $nm$ pulse to eliminate the residual 800 $nm$ light. Polarizations of the 800 $nm$ and 400 $nm$ fields were independently rotated with half-waveplates. Polarization angles of the 800 $nm$ and 400 $nm$ fields were set at +$48^{\circ}$ and -$64^{\circ}$ with respect to the maximum emission angle, which we define as the tip axis. We interpreted this axis to coincide with the crystalline facet normal. The two polarization angles were chosen to keep the total electron count rates below the repetition rate of the amplified laser (1 kHz). A translation stage with a piezo-transducer and a micrometer was used to control the temporal overlap of the two pulses. Using FROG and frequency-summing, the pulse duration of the 800 $nm$ and 400 $nm$ pulses were measured to be approximately $1\times10^2$ fs and $4\times10^2$ fs respectively \cite{Trebino}. In the vacuum chamber, a gold-coated off-axis parabolic mirror focused the 800 $nm$ and 400 $nm$ beams to a spot size of 7.8 $\mu m$ and 5.5 $\mu m$ diameter respectively. The nanotip was negatively biased at -170 $V$ without DC emission. The DC field strength at the tip apex is estimated to be $E_{dc} = 8.5 \times 10^{8}$ $V/m$, using $E_{dc} = V_{dc}/kr$ with tip voltage $V_{dc} = 170$ $V$, tip radius $r = 50$ $nm$, and field enhancement factor $k = 4$ \cite{Kruger2012}. A neutral density filter was placed in the optical arm of the 800 $nm$ pulse to reduce its power. Peak intensities of the 800 $nm$ and 400 $nm$ pulses were estimated to be $6.7 \times 10^{11}$ $W/cm^{2}$ (solid triangle in \Fig{fig:intensity}) and $2.2 \times 10^{10}$ $W/cm^{2}$ (solid square in \Fig{fig:intensity}) at the focus, respectively. The relative intensity values $I_{\w}/I_{2\w}$ were chosen so that signals from two-color and single-color multiphoton emission had comparable strength. Photoelectrons were collected with a channeltron detector and recorded by a counter with a 10-second average. 

\section{Results}

The intensity dependence of single-color photoemission was recorded. Laser parameters were set in the weak-field regime, so that multiphoton emission was dominant over tunneling photoemission \cite{Kruger2012}. In \Fig{fig:intensity}, the linear slopes confirm the multiphoton nature of the photoemission processes. For an 800 $nm$ field, the slope of $n=4$ indicates a four-photon emission process. For a 400 $nm$ field, the slope of $n = 2$ indicates a two-photon emission process. The inferred work function of the tungsten nanotip is between 4.5 eV and 6 eV, consistent with previously reported values \cite{Barwick}. The high work function can be accounted for by low electron emitting facets such as the W(011) crystalline plane \cite{Muller}. Assuming a nominal work function of 6 eV, the Schottky effect gives an effective work function of 4.9 eV \cite{Kruger2012}. The signature of multiphoton emission motivates the use of high-order time-dependent perturbation theory. The emission probabilities through the 800 $nm$ and 400 $nm$ single-color multiphoton channels are $P_{\w} = |C^{(4)}_{\w}|^2$ and $P_{2\w} = |C^{(2)}_{2\w}|^2$, where the corresponding probability amplitudes are
\begin{equation}\label{single-color}
	\begin{split}
	C^{(4)}_{\w} &=  \sum_{m,n,k} \int_{-\infty}^{\infty} dt_1 \bra{f} \hat{H}_{\w} (t_1) \ket{m}  \int_{-\infty}^{t_1} dt_2 \bra{m} \hat{H}_{\w} (t_2) \ket{n}	\\
		     & \times \int_{-\infty}^{t_2} dt_3 \bra{n} \hat{H}_{\w}(t_3) \ket{k} \int_{-\infty}^{t_3} dt_4 \bra{k} \hat{H}_{\w} (t_4) \ket{i},	\\
	C^{(2)}_{2\w} &= \sum_{m} \int_{-\infty}^{\infty} dt_1 \bra{f} \hat{H}_{2\w} (t_1) \ket{m}  \int_{-\infty}^{t_1} dt_2 \bra{m} \hat{H}_{2\w} (t_2) \ket{i} .
	\end{split}
\end{equation}
Here $\w$ and $2\w$ denote the 800 $nm$ and 400 $nm$ fields respectively. Notations $\ket{f}$ and $\ket{i}$ represent initial and final states, while $\ket{m}$, $\ket{n}$, and $\ket{k}$ are intermediate states. Summations are over all virtual transitions. As the tip size is small compared to laser wavelengths, the dipole approximation is assumed. The interaction Hamiltonian is taken to be 
\begin{equation}\label{Hamiltonian}
	\hat{H}_{\w, 2\w} (t) = -\hat{d} {E}^{(x)}_{\w, 2\w}(t),
\end{equation}
where $\hat{d}$ is the dipole operator, and ${E}^{(x)}_{\w, 2\w}(t)$ is the projected field along the tip axis. From \Eq{single-color}, it can be derived that the 800 $nm$ and 400 $nm$ single-color photoemission scale with field intensities and polarization angles as 
\begin{equation}\label{single-color_probw}
	P_{\w} \propto I^{4}_{\w}\cos^{8}{(\theta_{\w})}	,
\end{equation} 
\begin{equation}\label{single-color_prob2w}
	P_{2\w} \propto I^{2}_{2\w}\cos^{4}{(\theta_{2\w})},
\end{equation} 
where $I_{\w}$ ($I_{2\w}$) and $\theta_{\w}$ ($\theta_{2\w}$) are the field intensity and polarization angle of the 800 $nm$ (400 $nm$) pulse. When the 800 $nm$ and 400 $nm$ pulses overlap, the background-subtracted data shows a linear slope of $n=3$ (See \Fig{fig:intensity}), evidencing the presence of a multicolor quantum channel. The three-photon two-color quantum channel has the emission probability amplitude
\begin{equation}\label{multicolor}
	\begin{split}
	C^{(3)}_{\w+2\w} & = \sum_{p(\w,2\w)} \sum_{m,n} \int_{-\infty}^{\infty} dt_1 \bra{f} \hat{H}_{\w} (t_1-\tau_{\w}) \ket{m}  	\\
			     & \times \int_{-\infty}^{t_1} dt_2 \bra{m} \hat{H}_{\w} (t_2-\tau_{\w}) \ket{n}  \int_{-\infty}^{t_2} dt_3 \bra{n} \hat{H}_{2\w}(t_3) \ket{i},
	\end{split}
\end{equation}
where $\tau_{\w}$ is the delay of the $\w$-pulse. Summations are over all virtual transitions and permutations $p(\w,2\w)$ between $\hat{H}_{\w}$ and $\hat{H}_{2\w}$. The emission probability through the multicolor quantum channel is $P_{\w+2\w} = |C^{(3)}_{\w+2\w}|^2$, and it scales with field intensities and polarization angles as
\begin{equation}\label{multicolor_prob}
	P_{\w+2\w} \propto I^{2}_{\w}I_{2\w}\cos^{4}{(\theta_{\w})}\cos^{2}{(\theta_{2\w})}. 
\end{equation}
One important remark is that multiphoton emission depends on field intensities, while tunneling emission depends on the fields themselves. This is because photoemission in the weak fields is accomplished through multiphoton transitions, as described by perturbative amplitudes which depend on the time integral of the interaction, rather than the instantaneously modified work function as used in the description of tunneling photoemission. As a result, multiphoton emission depends on only the polarization of both fields with respect to the tip axis, while tunneling emission also depends on the relative polarization between the fields.

\begin{figure}[t]
\centering
\scalebox{0.35}{\includegraphics{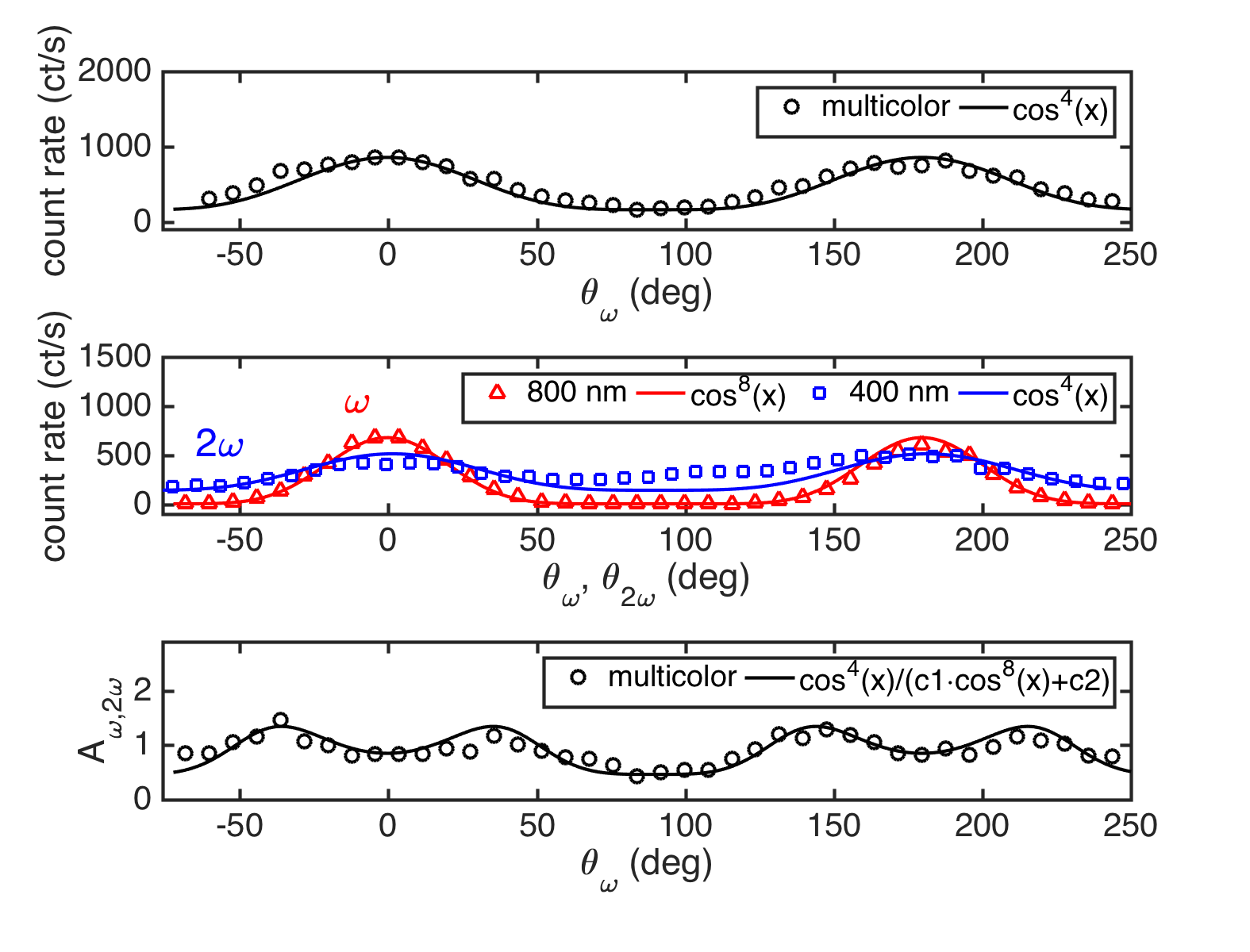}}
\caption{Polarization dependence of single-color and two-color multiphoton emission. Top: the background-subtracted two-color signal (black circle) shows a $\cos^4{(\theta_{w})}$ polarization dependence, implying that the multicolor quantum channel involves two $\w$-photons. Middle: the 800 $nm$ and 400 $nm$ signals scale with four-photon polarization dependence $\cos^8{(\theta_{\w})}$ and two-photon polarization dependence $\cos^4{(\theta_{2\w})}$, respectively. Bottom: the double maxima in additivity $A_{\w,2\w}$ confirms that fewer $\w$-photons are absorbed in the multicolor quantum channel than in the 800 $nm$ single-color channel.} 
\label{fig:polarization}
\end{figure}

Polarization measurements support this picture of two-color multiphoton emission. Varying $\theta_{\w}$ while keeping $\theta_{2\w} = -64^{\circ}$ fixed, makes the two-color photoemission vary with $\cos^{4}{(\theta_{\w})}$, as shown in the top panel of \Fig{fig:polarization}. This agrees with \Eq{multicolor_prob}. The zero polarization angles, $\theta_{\w} = 0$ and $\theta_{2\w} = 0$, are aligned with the tip axis. Unlike two-color tunneling photoemission, the relative polarization angle $\theta_{\w} - \theta_{2\w}$ does not play a significant role here (See \Eq{multicolor_prob}). This is clearly indicated by the data, where the two-color signal is symmetric with respect to the tip axis ($\theta_{\w} = 0^{\circ}$) instead of the fixed 400 $nm$ polarization angle ($\theta_{2\w} = -64^{\circ}$). In the middle panel of \Fig{fig:polarization}, red triangles (blue squares) gives the single-color polarization dependence, which is obtained by sending in only the 800 $nm$ (400 $nm$) pulse and varying $\theta_{\w}$ ($\theta_{2\w}$). This agrees with \Eq{single-color_probw} (\Eq{single-color_prob2w}) and shows that the two-color signal has a broader polarization width than the 800 $nm$ single-color signal (cf. top panel black curve and middle panel red curve of \Fig{fig:polarization}). This is because the two $\w$-photons absorbed in the multicolor channel give a polarization dependence of $\cos^{4}{(\theta_{\w})}$, while the four $\w$-photons absorbed in the 800 $nm$ single-color channel give a dependence of $\cos^{8}{(\theta_{\w})}$. 

In the bottom panel of \Fig{fig:polarization}, the additivity 
\begin{equation}\label{additivity}
	A_{\w,2\w} \equiv \frac{P_{\w,2\w}-P_{\w}-P_{2\w}}{P_{\w}+P_{2\w}}
\end{equation}
is given as a function of $\theta_{\w}$. Additivity is a convenient measure for collaborative effects in nonlinear systems \cite{Barwick}. A value of zero means that the system behaves in a linear fashion. Deviation from zero for a nonlinear system indicates the presence of collaborative effects. Additivity also characterizes quantum efficiency. $A_{\w,2\w} = 1$ corresponds to a twofold increase in quantum efficiency. Substituting $P_{\w,2\w} = P_{\w} + P_{2\w} + P_{\w+2\w}$ to \Eq{additivity} and using \Eqsx{single-color_probw}{single-color_prob2w}{multicolor_prob}, gives the additivity 
\begin{equation}
	A_{\w,2\w} = \frac{P_{\w+2\w}}{P_{\w} + P_{2\w}} = \frac{\cos^{4}{(\theta_{\w})}}{c_1\cos^{8}{(\theta_{\w})}+c_2}	,
\end{equation}
where $c_1$ and $c_2$ are parameters controlled by the field intensities. The additivity shows a double maxima because the single-color signal in the denominator is narrower than the two-color signal in the numerator.

\begin{figure}[t]
\centering
\scalebox{0.4}{\includegraphics{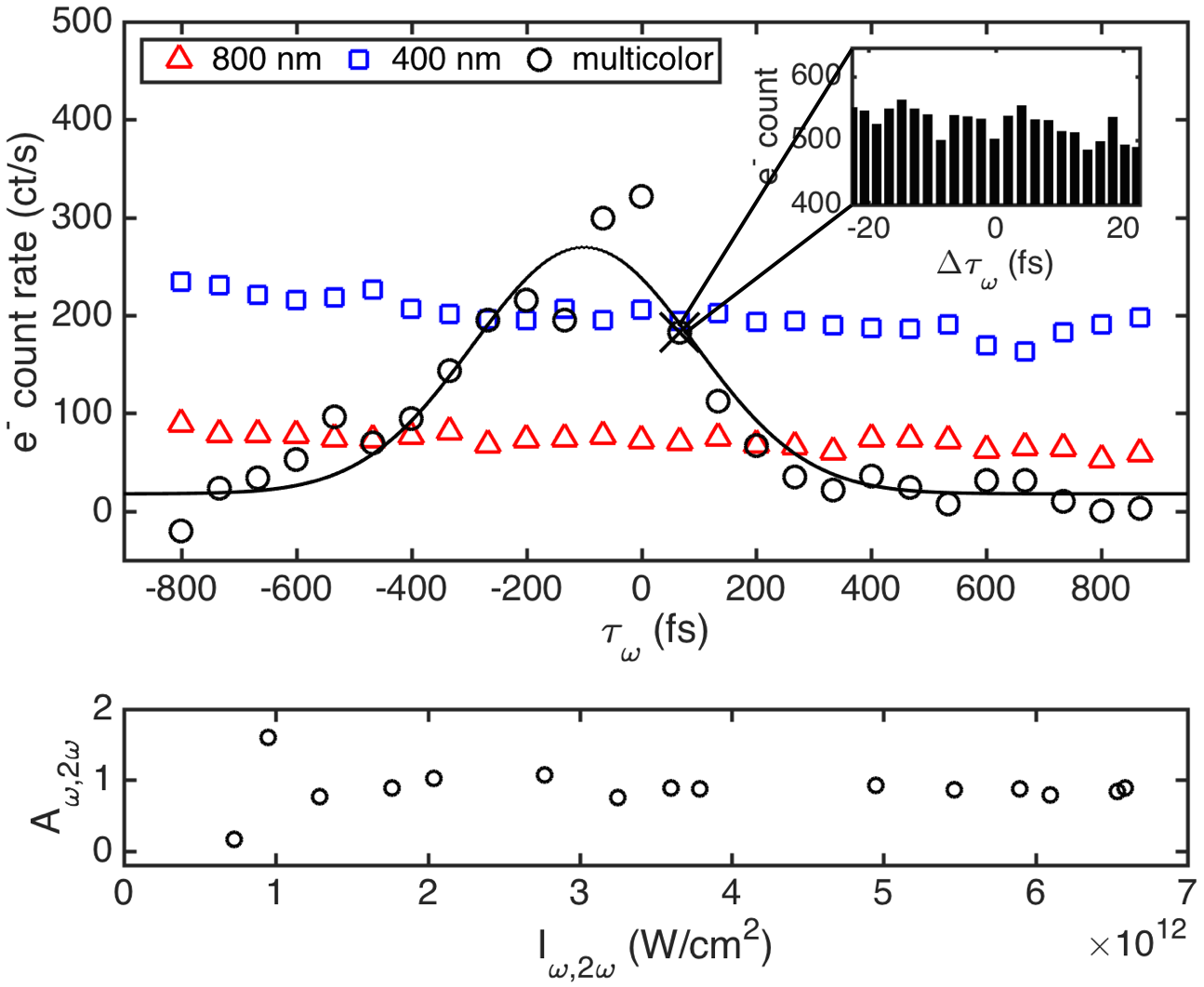}}
\caption{Electron correlation spectrum of two-color multiphoton emission. Top: the overlap of the two-color pulses opens the multicolor quantum channel (black circle). Simulation based on two-color multiphoton emission gives good agreement with the data (black line). Top inset: no fringes were observed in the electron correlation spectrum. The translation stage was parked at $\tau_{\w} = 6.7 \times 10$ fs (black cross), while the piezo-transducer scanned through a delay range of 40 fs. Bottom: the twofold increase in quantum efficiency is stable with increasing two-color field intensity $I_{\w,2\w}$.}
\label{fig:delay}
\end{figure}

Time-delay measurements show the opening of the multicolor quantum channel. In the top panel of \Fig{fig:delay}, the time-delay electron correlation spectrum shows a clear peak that is due to two-color multiphoton emission. Numerically solving the Schr\"{o}dinger equation using the Hamiltonian in \Eq{Hamiltonian} and a 2-level model, gives good agreement with the data. In the bottom panel of \Fig{fig:delay}, the twofold increase in quantum efficiency is shown to be stable with increasing two-color field intensity $I_{\w,2\w} \equiv (I^2_{\w}I_{2\w})^{1/3}$. 

Notably, no fringes were observed in the electron correlation spectrum (inset of top panel, \Fig{fig:delay}), implying that the multicolor quantum channel is controlled by the pulse delay but not the relative phase between the two fields. In contrast to two-color tunneling photoemission, phase effects in two-color multiphoton emission occur only if quantum interference is allowed. This requires that identical initial and final states can be reached through multiple quantum channels. The fact that we do not observe fringes in the electron correlation spectrum suggests that the final states reached by the multiphoton channels are not the same. 

During the review process of this paper, a similar work from Hommelhoff's group was published, where they observed interference fringes in the electron correlation spectrum \cite{Forster}. In their experiment, a single crystalline W(310) nanotip with an effective work function 3.6 eV was irradiated with 1560 $nm$ and 780 $nm$ femtosecond pulses. The observed fringes are attributed to the presence of a strong intermediate state right below the effective work function. The intermediate state facilitates quantum interference between the four-photon 1560 $nm$ channel and the three-photon multicolor channel. As polycrystalline tungsten nanotips were used in our experiment, it is likely that there was no such an intermediate state, thus no interference was possible. Nevertheless, it is noteworthy that the model developed through our experiment is able to predict the shape of observed fringe pattern and the periodicity in Hommelhoff's experiment, for which they stated ``... fail to describe the sinusoidal shape observed in the experiment'' \cite{Forster}. 

The experimental parameters they used were such that photoemission from the two-photon 780 $nm$ channel is negligible, so interference occurs between only two multiphoton channels. For the four-photon 1560 $nm$ channel, the emission probability amplitude is similar to \Eq{single-color},
\begin{equation}\label{W310_single}
	\begin{split}
	C^{(4)}_{\nu} &=  \sum_{m,n,k} \int_{-\infty}^{\infty} dt_1 \bra{f} \hat{H}_{\nu} (t_1-\tau_{\nu}) \ket{m}  \int_{-\infty}^{t_1} dt_2 \bra{m} \hat{H}_{\nu} (t_2-\tau_{\nu}) \ket{n}	\\
		     & \times \int_{-\infty}^{t_2} dt_3 \bra{n} \hat{H}_{\nu}(t_3-\tau_{\nu}) \ket{k} \int_{-\infty}^{t_3} dt_4 \bra{k} \hat{H}_{\nu} (t_4-\tau_{\nu}) \ket{i},	
	\end{split}
\end{equation}
where $\nu$ stands for the 1560 $nm$ field, and $\tau_{\nu}$ is the delay of the $\nu$-pluse. The accumulated phase factor is $\exp{(+ i 4\nu\tau_{\nu})}$ because four $\tau_{\nu}$-dependent factors are involved. The positive sign indicates the absorption of $\nu$-photons. For the three-photon multicolor channel, the emission probability amplitude is  similar to \Eq{multicolor},
\begin{equation}\label{W310_multicolor}
	\begin{split}
	C^{(3)}_{\nu+2\nu} & = \sum_{p(\nu,2\nu)} \sum_{m,n} \int_{-\infty}^{\infty} dt_1 \bra{f} \hat{H}_{\nu} (t_1-\tau_{\nu}) \ket{m}  	\\
			     & \times \int_{-\infty}^{t_1} dt_2 \bra{m} \hat{H}_{\nu} (t_2-\tau_{\nu}) \ket{n}  \int_{-\infty}^{t_2} dt_3 \bra{n} \hat{H}_{2\nu}(t_3) \ket{i},
	\end{split}
\end{equation}
where $2\nu$ denotes the 780 $nm$ field. The accumulated phase factor is $\exp{(+ i 2\nu\tau_{\w})}$ due to the absorption of two $\nu$-photons. Using $C^{(4)}_{\nu} = f_{\nu}\exp{(+ i 4\nu\tau_{\nu})}$ and $C^{(3)}_{\nu+2\nu} = f_{\nu+2\nu}g(\tau_{\nu})\exp{(+ i 2\nu\tau_{\nu})}$, where $f_{\nu} \propto I^2_{\nu}\cos^4{(\theta_{\nu})}$ and $f_{\nu+2\nu} \propto I_{\nu}\sqrt{I_{2\nu}}\cos^2{(\theta_{\nu})}\cos{(\theta_{2\nu})}$ are real-valued constants, and $g(\tau_{\nu})$ is a normalized real-valued convolution function, it can be shown that the interference fringes follow a sinusoidal pattern 
\begin{equation}\label{fringe}
	\begin{split}
	P(\tau_{\nu}) &= |C^{(4)}_{\nu}|^2 + |C^{(3)}_{\nu+2\nu}|^2 + C^{(4)\ast}_{\nu}C^{(3)}_{\nu+2\nu} + C^{(4)}_{\nu}C^{(3)\ast}_{\nu+2\nu}  	\\
				     &= f^2_{\nu} + f^2_{\nu+2\nu}g^2(\tau_{\nu}) + 2\cos{(\Delta \varphi_{qm})}f_{\nu}f_{\nu+2\nu}g(\tau_{\nu}),
	\end{split}
\end{equation}
where the phase difference between the two interfering quantum channels is $\Delta \varphi_{qm} = 2\nu\tau_{\nu}$. This result remains the same if the $2\nu$-pulse is delayed instead. The fringe periodicity from the above equation is $T = 2\pi/2\nu = 2.6$ fs, which corresponds to an oscillation frequency of 385 THz. The single-color signal $f^2_{\nu}$ gives an offset to the electron correlation spectrum. The multicolor signal $f^2_{\nu+2\nu}g^2(\tau_{\nu})$ gives a peak similar to that observed in our experiment (See \Fig{fig:delay}). Visibility of the fringes is determined by the ratio between the interference signal and the sum of the single-color and multicolor signals, 
\begin{equation}
	\mathcal{V} = \frac{2f_{\nu}f_{\nu+2\nu}}{f^2_{\nu} + f^2_{\nu+2\nu}} = \frac{I^3_{\nu}\sqrt{I_{2\nu}}\cos^6{(\theta_{\nu})}\cos{(\theta_{2\nu})}}{a_{1}I^4_{\nu}\cos^8{(\theta_{\nu})} + a_{2}I^2_{\nu}I_{2\nu}\cos^4{(\theta_{\nu})}\cos^2{(\theta_{2\nu})}}, 
\end{equation}
where $a_{1}$ and $a_{2}$ are parameters determined by material properties. The above analysis provides a physically motivated model that accounts for both ours and Hommelhoff's results, including the sinusoidal fringe shape, the 390 THz fringe oscillation frequency, and the power dependence of the fringe visibility.

\section{Discussion: Ultrafast Spin-Polarized Electron Sources}

The observed two-color multiphoton emission shows that multicolor quantum channels can be of comparable strength to single-color quantum channels. This provides the basis for realizing an ultrafast spin-polarized electron source using two-color multiphoton emission from a nanotip. An ultrafast spin-polarized nanostructured electron source is important for ultrafast electron microscopy and ultrafast electron diffraction \cite{Barwick2009, Feist, Siwick, Gulde}. The state-of-the-art spin-polarized electron source is based on a NEA-GaAs photocathode, which is not ultrafast \cite{Pierce, Dreiling2014, Kuwahara}. Using two-color pulses, Sipe and colleagues have demonstrated ultrafast control for optically injected spin currents on semiconductor surfaces \cite{Bhat, Hubner, Stevens}. In ZnSe, two single-color quantum channels were interfered to create a net spin current \cite{Hubner}; one channel is 400 $nm$ one-photon excitation from valence to conduction band, and the other is 800 $nm$ two-photon excitation. We envision that such a technique can be used for spin current injection at the apex of a semiconductor nanotip, followed by extraction of spin-polarized electrons via two-color multiphoton photoemission \cite{Kuwahara}. The nanotip allows the use of low-intensity fields, as compared to surface emission, and leads to a spatially coherent electron source \cite{Cho, Ehberger}. Although our experiment did not show interference effects, the demonstrated multicolor quantum channel and its control may be used for launching and extracting ultrafast spin-polarized photoelectrons in appropriate materials.
  
\section{Conclusion}

In conclusion, two-color multiphoton emission from a tunsten nanotip has been demonstrated. The two-color multiphoton emission is assisted by a three-photon multicolor quantum channel. The multicolor channel led to a twofold increase in quantum efficiency. Control of two-color multiphoton emission was achieved by opening and closing the multicolor quantum channel with pulse delay. The demonstrated two-color multiphoton emission provides a pathway for the possible realization of ultrafast spin-polarized electron sources via optically injected spin current.  

\section{Acknowledgements}

The authors thank Cornelis J. Uiterwaal, Timothy J. Gay, Sy-Hwang Liou, and Yen-Fu Liu for advice. We thank Eric Jones for laser focal size measurements and suggestions of spin-polarized electron sources via spin-currents. W. C. Huang wishes to thank Yanshuo Li for insightful discussions. This work utilized high-performance computing resources from the Holland Computing Center of the University of Nebraska. Manufacturing and characterization analyses were performed at the NanoEngineering Research Core Facility (part of the Nebraska Nanoscale Facility), which is partially funded from the Nebraska Research Initiative. Funding for this work comes from National Science Foundation (NSF) grants EPS-1430519 and PHY-1306565.

\end{document}